\documentstyle[psfig,aps,twocolumn,pre]{revtex} 

\def \eurpL#1#2#3#4{{ Europhys.~Lett. {\bf #1}, #3 (#2)}}

\def \jcg#1#2#3#4{{ J.~Cryst.~Growth {\bf #1}, #3 (#2)}}
\def \jcompp#1#2#3#4{{ J.~Comp.~Phys. {\bf #1}, #3 (#2)}}

\def \jpa#1#2#3#4{{ J.~Phys.~A: Math.~Gen. {\bf #1}, #3 (#2)}}
\def \jpc#1#2#3#4{{ J.~Phys.~C: Solid State {\bf #1}, #3 (#2)}}

\def \jpq1#1#2#3#4{{ J.~Physique {\it I} (France) {\bf #1}, #3 (#2)}}

\def \jsp#1#2#3#4{{  J.~Stat.~Phys. {\bf #1}, #3 (#2)}}
\def \lanl#1{{\tt (cond-mat/#1)}}

\def \pha#1#2#3#4{{ Physica A {\bf #1}, #3 (#2)}}

\def \prb#1#2#3#4{{ Phys.~Rev.~B {\bf #1}, #3 (#2)}}
\def \pre#1#2#3#4{{ Phys.~Rev.~E {\bf #1}, #3 (#2)}}
\def \prl#1#2#3#4{{ Phys.~Rev.~Lett. {\bf #1}, #3 (#2)}}
\def \repp#1#2#3#4{{ Rep.~Prog.~Phys. {\bf #1}, #3 (#2)}}

\def \siamjc#1#2#3#4{{ SIAM J.~Comput. {\bf #1}, #3 (#2)}}

\def \zphysb#1#2#3#4{{ Z.~Physik. {\bf #1}, #3 (#2)}}

\newcommand{\RP}{FeSe,KaWe,Tho1,Ber,DaTr,ThGa,RH0,ArSah0,RH2,KnSah,ArSah,Obkv,Lett1,JaTh1,JaTh2,Lett2,Cmt,Cayley,FM,Cmp2,Cmp}
\newcommand{\MFTheories}{Obkv,Lett2,Cayley}
\newcommand{\Algorithms}{Hen,CFMalg,PG,Ja}
\newcommand{\Numerical}{Lett1,JaTh1,JaTh2,Lett2,Cmt,FM,Cmp2,Cmp}
\newcommand{\BP}{CLR,AL,Duarte,Sch,VE,AdSt,Adler,ChaKo,MeCha}
\newcommand{\DP}{conjecture,Kinzel,Grass2+1,LSMJ,FHL,LFH,DK,BG,ET,EGJT}
\newcommand{\DPW}{LSMJ,FHL,LFH,ET,EGJT}

\begin{document}
\title{Directed rigidity and bootstrap percolation in (1$+$1) dimensions}
\author{Marcio Argollo de Menezes \footnote{marcio@if.uff.br} and Cristian
  F.~Moukarzel\footnote{cristian@if.uff.br}}
\address{Instituto de F\'\i sica, Universidade Federal Fluminense, \\
  CEP 24210-340, Niteroi RJ, Brazil.}  \maketitle
\begin{abstract}
  We study directed rigidity percolation (equivalent to directed bootstrap
  percolation) on three different lattices: square, triangular, and augmented
  triangular. The first two of these display a first-order transition at
  $p=1$, while the augmented triangular lattice shows a continuous transition
  at a non-trivial $p_c$.  On the augmented triangular lattice we find, by
  extensive numerical simulation, that the directed rigidity percolation
  transition belongs to the same universality class as directed percolation.
  The same conclusion is reached by studying its surface critical behavior,
  i.e. the spreading of rigidity from finite clusters close to a non-rigid
  wall. Near the discontinuous transition at $p=1$ on the triangular lattice,
  we are able to calculate the finite-size behavior of the density of rigid
  sites analytically. Our results are confirmed by numerical simulation.
\end{abstract}


\section{Introduction}
\label{sec:intro}
Central-force rigidity percolation (RP)~\cite{\RP} is the mechanical
equivalent of the usual percolation
problem~\cite{Sta-bk,Sah-bk,BuHa,Ess_rv,Sah-rv}. In RP forces (vectors) must
be transmitted instead of scalars.  This problem has received increased
attention recently, following the development of mean-field
theories~\cite{\MFTheories} as well as of powerful combinatorial
algorithms~\cite{\Algorithms} for its numerical study~\cite{\Numerical}. As a
result of these efforts, a deeper understanding of the rigidity transition has
emerged, although some open questions remain.

Bethe lattice calculations for RP~\cite{Lett2,Cayley} with an adjustable
number $g$ of degrees of freedom at each site have been used to obtain the
behavior of the spanning cluster density $P_{\infty}(p)$ as a function of $p$,
the dilution (bond or site) parameter. For $g=1$ one has usual (scalar)
percolation, displaying a continuous transition with $\beta^{MF}=1$. But for
any $g>1$, the order parameter $P_{\infty}$ has a discontinuity at a finite
critical value $p_c$. Thus the rigidity transition is discontinuous for $d\to
\infty$\cite{Cayley,FM}.  Other MF approximations also predict a first-order
RP transition~\cite{Obkv}

On triangular lattices on the other hand, there is a divergent correlation
length and the RP transition is \emph{second
  order}~\cite{Lett1,JaTh1,JaTh2,Lett2,Cmt,Cmp2,Cmp}, but in a different
universality class than usual percolation~\cite{Cmp2,Cmp}.  Some of the
numerical evidence in 2d is consistent with a small discontinuity in the order
parameter $P_{\infty}$, or a very small value for $\beta$, but the precise
interpretation of this evidence is still a matter of debate~\cite{Cmt}. In
three dimensions the rigidity transition is undoubtedly
second-order~\cite{CFMunpub}.  It is at present unclear in which fashion the
RP transition becomes discontinuous as the dimensionality increases.  Is there
something like an upper critical dimension for RP, beyond which it is
first-order? Or does it get increasingly ``first-order'' (i.e.  $\beta \to 0$)
as $d \to \infty$?  This analysis is further complicated by the fact that the
character of the transition is \emph{lattice} dependent. Hypercubic lattices
in which sites have $d$ degrees of freedom each cannot be rigid if they are
diluted, but are rigid if undiluted and if they have appropriate boundary
conditions.  Thus on hypercubic lattices the RP transition is ``trivially
first-order'' at $p_c=1$, in any dimension.

Similar considerations apply to directed lattices. Bethe lattices are directed
by construction, since there is only one path between any two given sites. On
directed lattices, rigid connectivity takes a particularly simple form.
Imagine a rigid boundary to which a site with $g$ degrees of freedom must be
rigidly attached by means of rotatable springs (central forces). Each spring,
or bond, restricts one degree of freedom. Thus the minimum number of bonds
required to completely fix this site is $g$. Propagation of rigidity on
directed lattices is then defined in the following terms: a site (with $g$
degrees of freedom) at ``time'' $t$ is rigidly connected to a boundary at
$t=0$ if and only if it has $g$ or more neighbors at \emph{earlier} times who
in turn are rigidly connected to the boundary. Thus, in contrast to
\emph{undirected} rigidity, which requires complex
algorithms~\cite{\Algorithms} that presently limit the maximum sizes to
approximately $1.6 \times 10^7$ sites~\cite{Cmp,Cmp2}, directed rigidity
percolation (DRP) can be studied by means of a simple numerical procedure, and
on much larger systems.

It is interesting to notice that, on any \emph{directed} lattice, rigidity
percolation is equivalent to \emph{bootstrap percolation} (BP), a modified
percolation problem in which a site belongs to a cluster if at least $m$ of
its neighbors also do~\cite{\BP}. Bootstrap percolation on undirected lattices
attempts to describe certain systems in which atoms behave magnetically only
if they are surrounded by a ``large enough'' number of magnetic neighbors. A
second reason for interest in BP is the search for novel critical behaviors in
percolation~\cite{AdSt,Adler}, but the present understanding of this problem
indicates that BP is either ``trivially first-order'' with $p_c=1$ or
second-order and in the universality class of scalar
percolation~\cite{ChaKo,MeCha,Nilton}.

Early studies of semi-directed $m=2$ BP on square lattices seemed to indicate
a transition at a non-trivial $p$~\cite{Duarte,VE}, but rigorous
arguments~\cite{Sch} later showed that $p_c=1$ in this case.  To our knowledge
there are no published studies of directed bootstrap percolation (DBP)
displaying a second-order transition.

It has been recently conjectured~\cite{conjecture} that any continuous
transition in a nonequilibrium process with a scalar order-parameter, and a
non-fluctuating, non-degenerate absorbing state must be in the same
universality class as directed percolation(DP)~\cite{\DP}. According to this,
DRP-DBP would belong to the DP universality class in all dimensions for which
it has a continuous transition.

It is thus interesting to study RP on finite-dimensional directed lattices of
increasing dimensionality, both to test this conjecture and to understand in
which fashion DRP becomes discontinuous as $d \to \infty$. In this article we
report on our results for directed rigidity percolation (DRP, equivalent to
DBP) on several (1$+$1)-dimensional lattices displaying first- and
second-order phase transitions.

\begin{figure}[htb] 
\centerline{ 
{\bf a}\hskip -0.3cm\psfig{figure=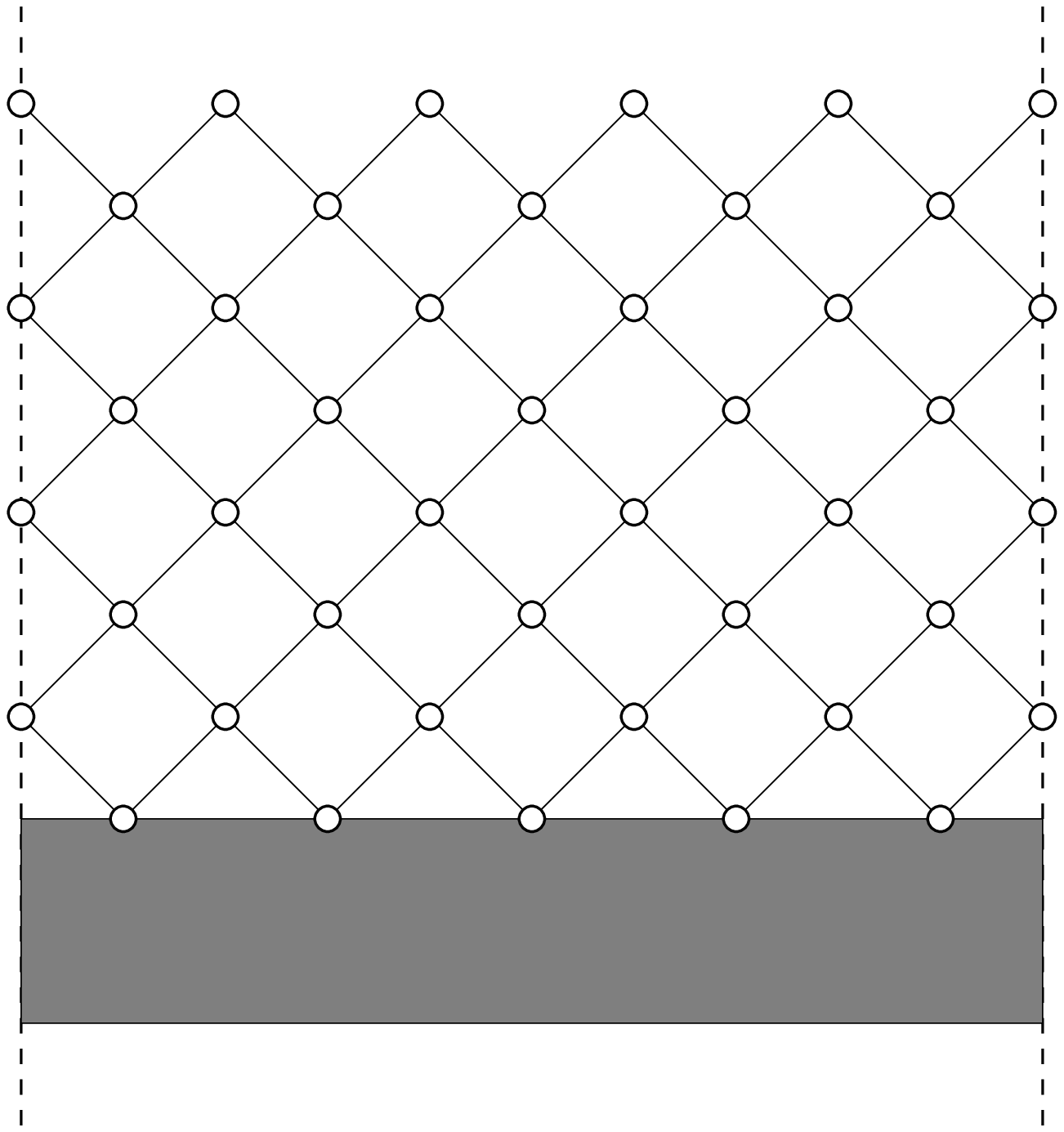,width=3cm,angle=270}
{\bf b}\hskip -0.3cm\psfig{figure=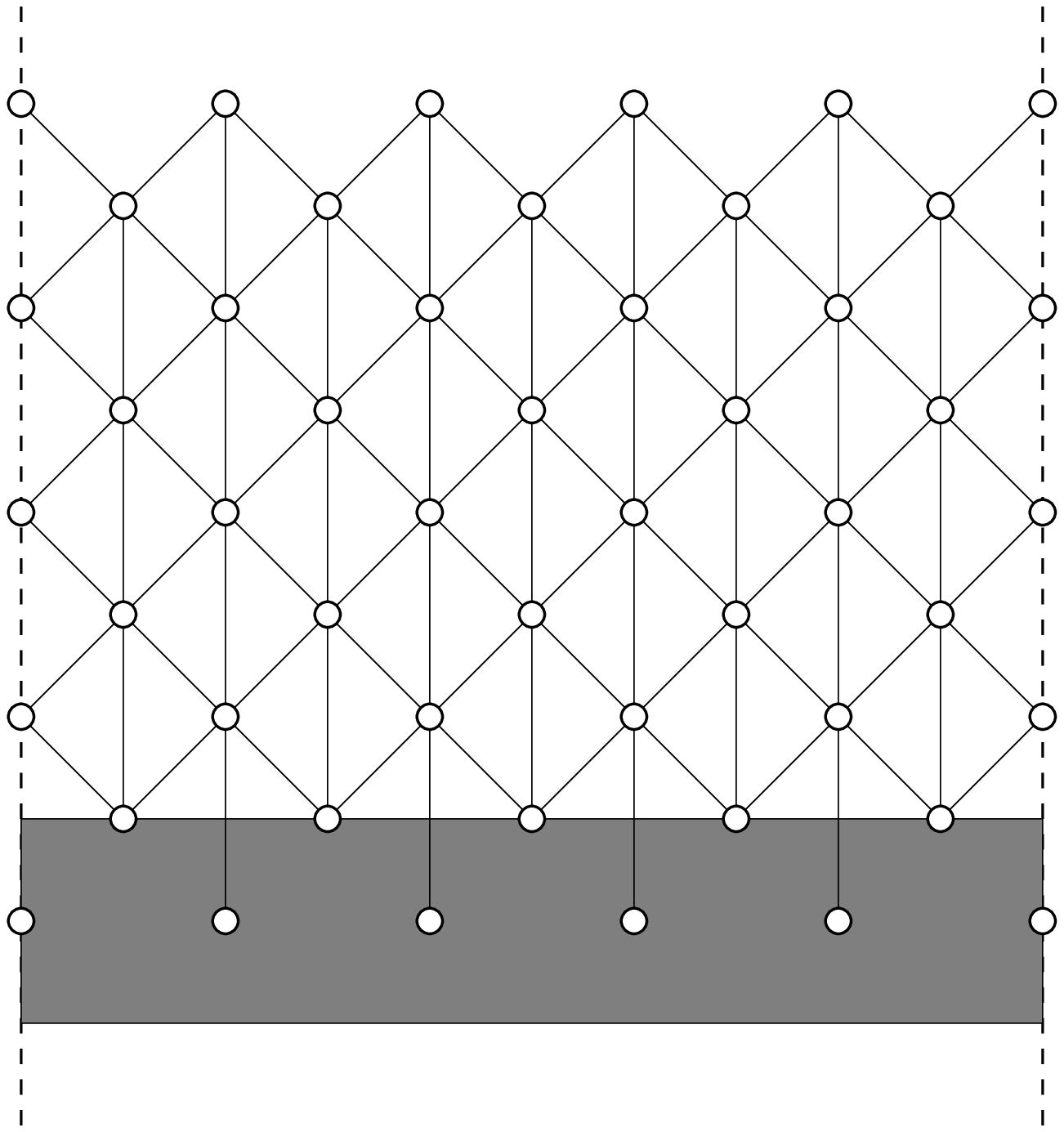,width=3cm,angle=270}
{\bf c}\hskip -0.3cm\psfig{figure=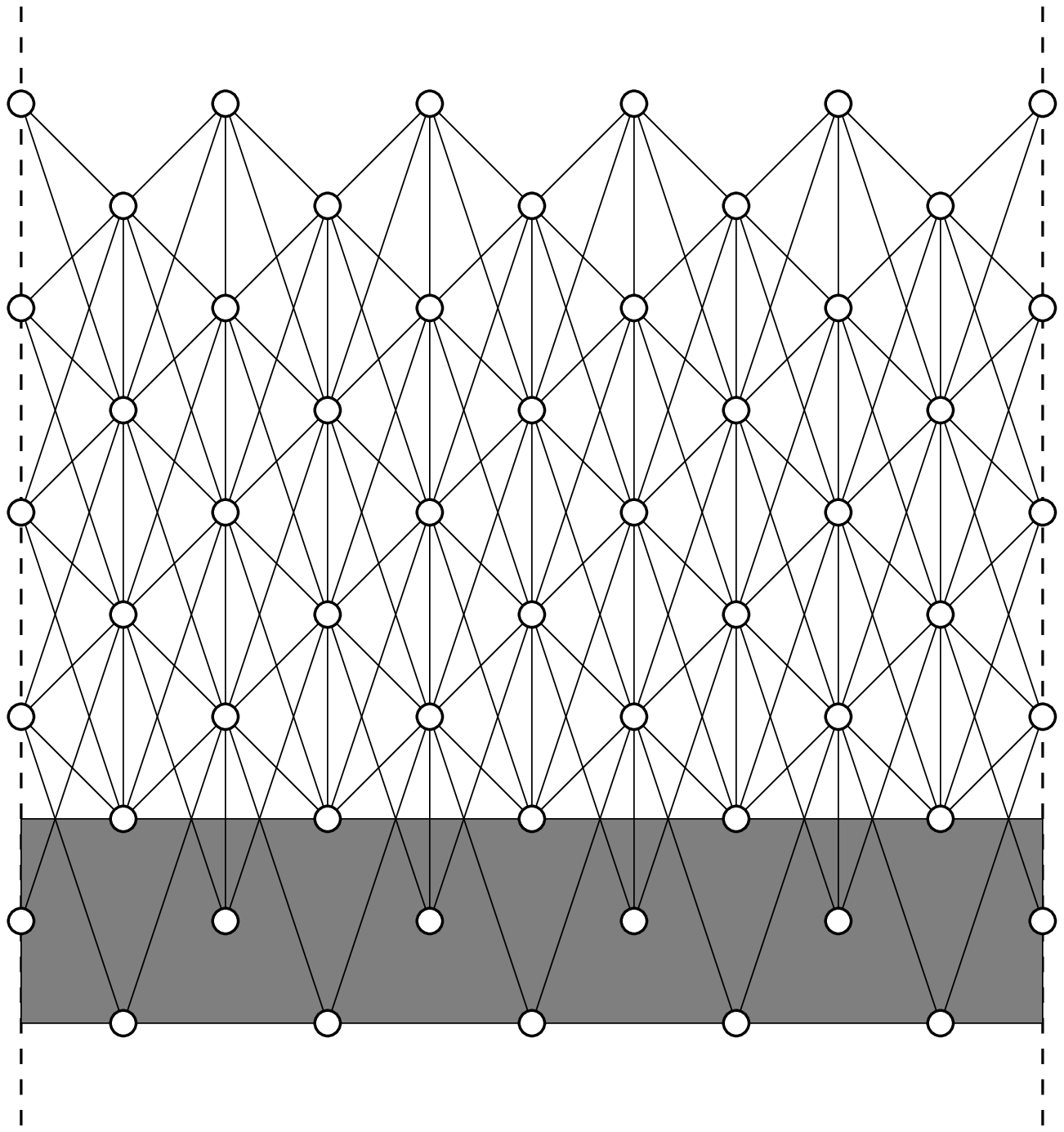,width=3cm,angle=270}
}
\vskip 0.5cm
\caption{{}
  Lattices used for directed rigidity percolation studies in this work. {\bf
    a)} square lattice, {\bf b)} triangular lattice, and {\bf c)}
  five-neighbor lattice.  All examples are shown here undiluted ($p=1$).
  Rigidity propagates upwards in ``time'' from the rigid boundary (gray) at
  $t=0$. A site is defined to be rigidly connected to this boundary if it has
  two or more neighbors at \emph{earlier} times which in turn are connected. }
\label{fig:latexample}
\end{figure}

If the DRP transition is second-order, we pay particular attention to the
determination of critical indices associated with the spreading of rigidity,
as described in the following~\cite{Grass2+1,LSMJ}.  As usual in the study of
directed processes, we define $D(p)$ to be the asymptotic density of
``active'' (rigid) sites, which is equivalent to the probability $P(p)$ that,
at large times $t$, a randomly chosen point be rigidly connected to a totally
rigid boundary at $t=0$.  If the dilution parameter $p$ is lower than a
critical value $p_c$, rigidity does not propagate and $P(p)=0$. If the
transition is second-order, immediately above $p_c$ one has $P(p) \sim
(p-p_c)^{\beta^{dens}}$.

If the evolution starts from a \emph{finite} rigid cluster or ``seed'' at
$t=0$ instead of a rigid boundary, one defines $P_a^{seed}(t,p)$ as the
probability that the cluster grown from this seed still be ``active'' at time
$t$.  If the transition is second-order, $P_a^{seed}(t,p) \sim
(p-p_c)^{\beta^{seed}}$ for $p \to p_c^+$ and $t\to \infty$. At $p_c$ this
quantity decays as
\begin{equation}
P_a^{seed}(t,p_c)  \sim t^{-\delta},
\label{eq:p-decay}
\end{equation}
with $\delta =\beta^{seed}/\nu_{\parallel}$ and $\nu_{\parallel}$ the temporal
(or parallel) \emph{correlation length} exponent: $\xi_{\parallel} \sim
|p-p_c|^{-\nu_{\parallel}}$

The typical width $w$ of a cluster grown from a finite seed at $p_c$ behaves
as
\begin{equation}
w(t) \sim t^{\chi},
\label{eq:x-growth}
\end{equation}
where $\chi=\nu_{\perp}/\nu_{\parallel}$ and $\nu_{\perp}$ is the critical
index associated with the decay length $\xi_{\perp}$ of perpendicular or
``space'' correlations: $\xi_{\perp} \sim |p-p_c|^{-\nu_{\perp}}$. Averages
are taken only over clusters still alive at time $t$. Finally, the average
mass of a cluster grown from a finite seed at $p_c$ behaves as
\begin{equation}
M_{seed}(t) \sim t^{\tilde{\eta}},
\label{eq:m-growth}
\end{equation}
where $\tilde{\eta} = (\nu_{\parallel} + \nu_{\perp} - \beta^{dens}) /
\nu_{\parallel}$~\cite{LSMJ}.

For comparison we also simulate numerically usual directed percolation (DP,
which corresponds to $g=1$). In the DP case a simple argument shows that
$\beta^{dens}=\beta^{seed}$ because of time-reversal symmetry: consider for
simplicity bond dilution and choose an arbitrary point $x$ at time $t$. Any
configuration of occupied bonds connecting $(x,t)$ to the boundary at $t=0$,
and thus contributing to $P(p,t)$, when reflected in the time direction,
contributes to $P_a^{seed}(t,p)$ if now a point-like seed is located at $x$.
Since both the original and the time-inverted configuration have the same
probability, $P(p,t)=P_a^{seed}(p,t)$ exactly for bond-diluted DP and
therefore $\beta^{dens}=\beta^{seed}$. Notice that this equality implies
$\tilde{\eta} + \delta - \chi =1$.

\begin{figure}[htb] 
\centerline{ 
{\bf a}\hskip -0.3cm
\psfig{figure=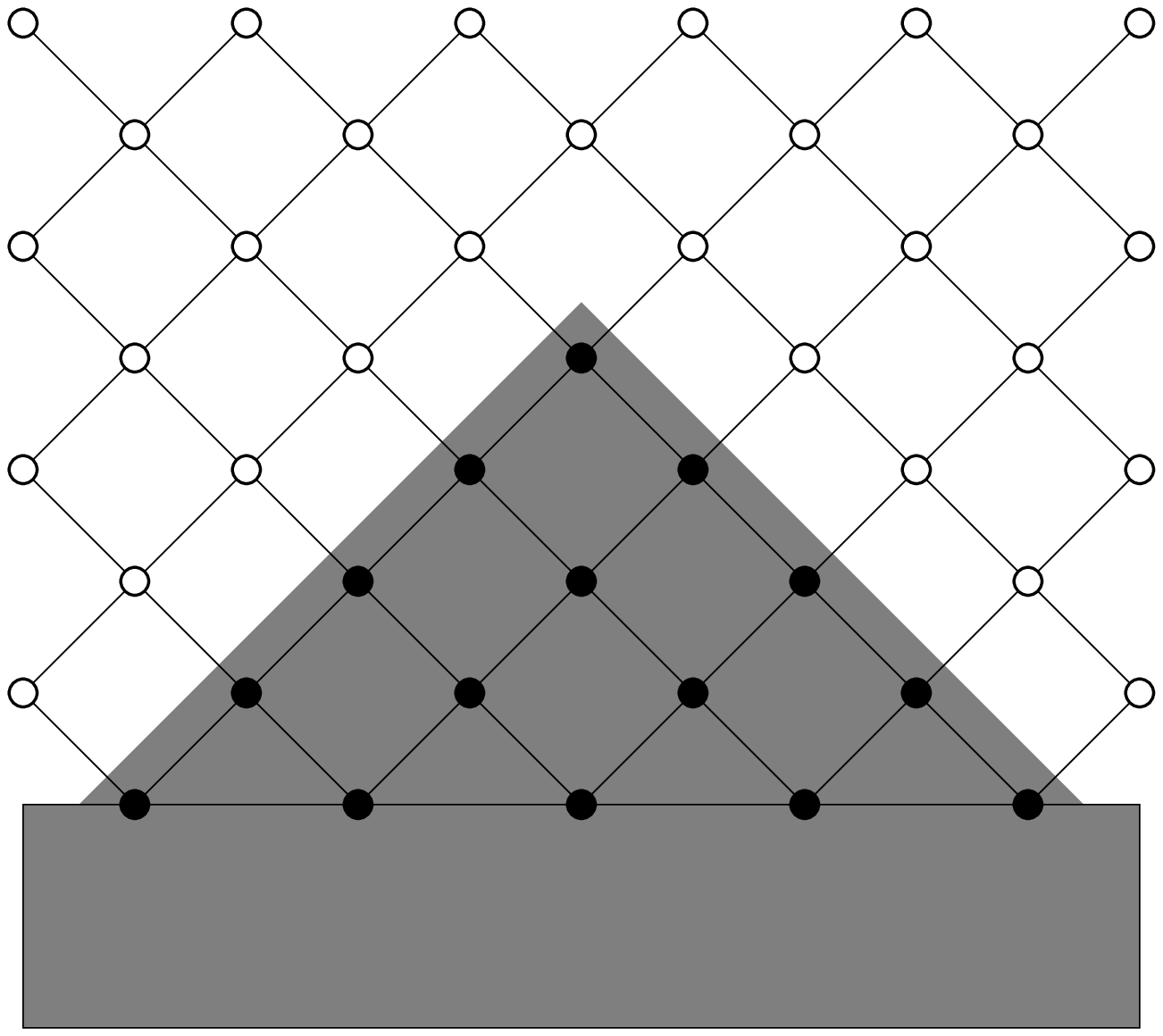,width=3cm,angle=270}
{\bf b}\hskip -0.3cm
\psfig{figure=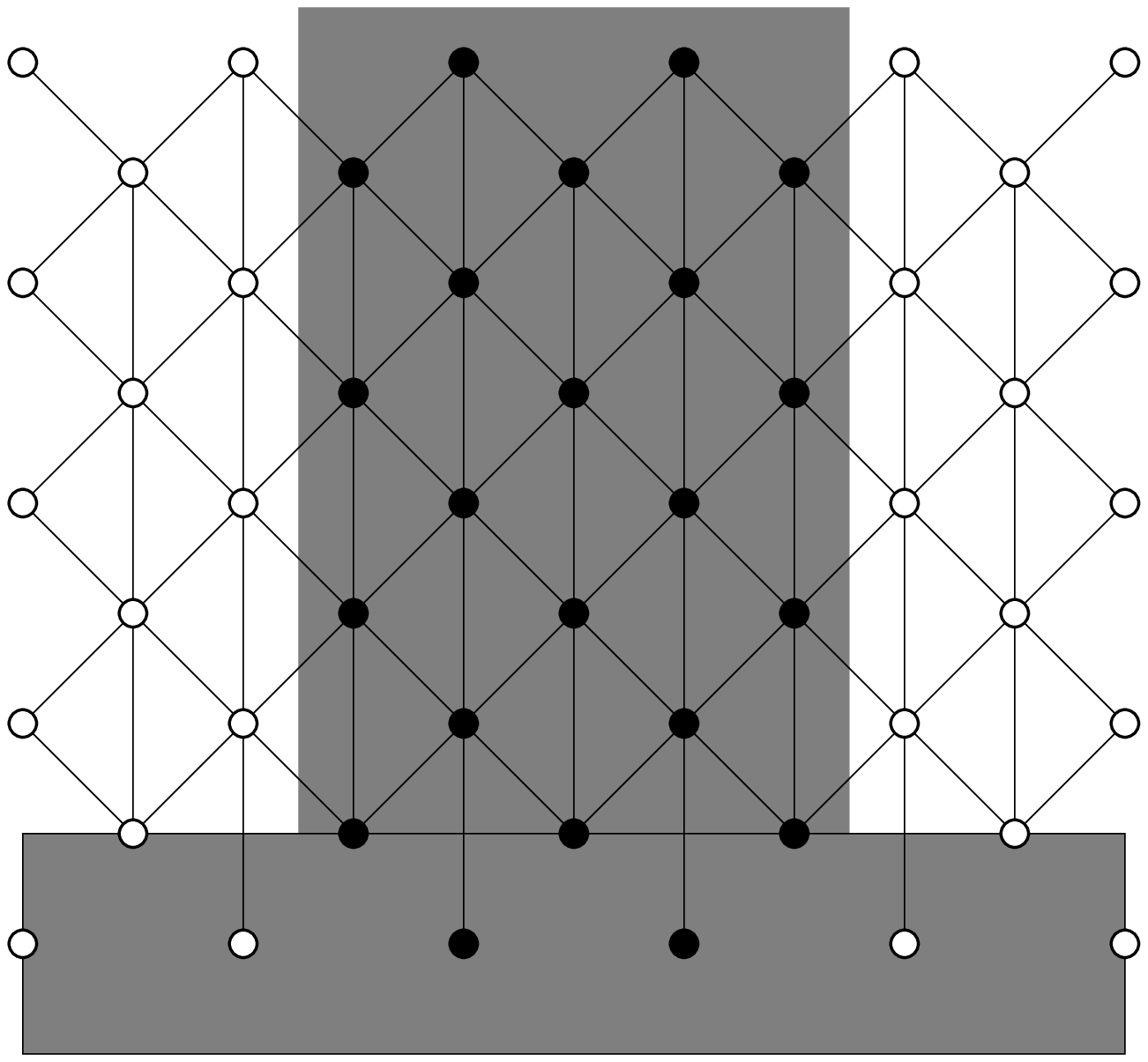,width=3cm,angle=270}
{\bf c}\hskip -0.3cm
\psfig{figure=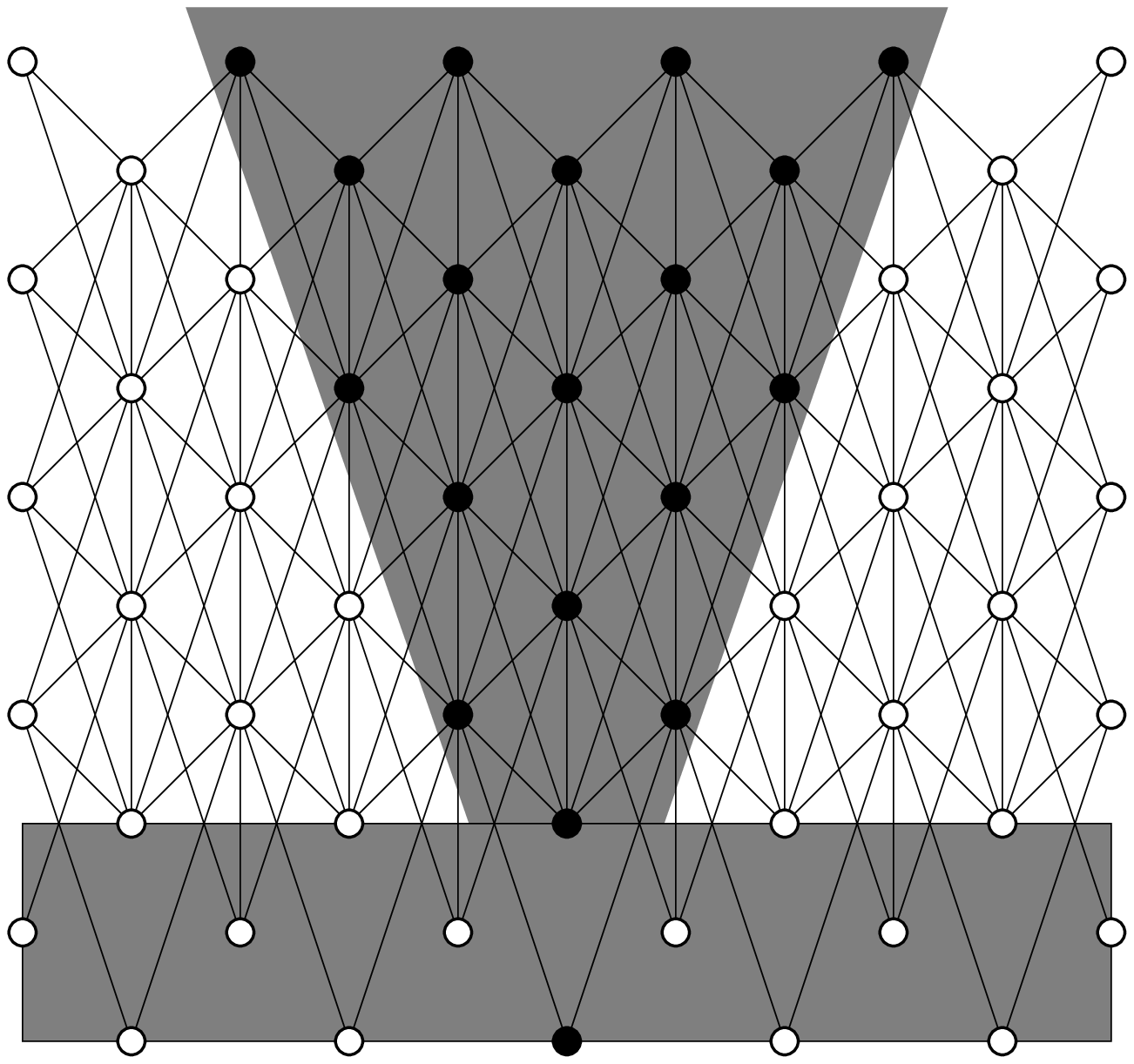,width=3cm,angle=270}
}
\vskip 0.5cm
\caption{{} Propagation of rigidity from finite clusters. The starting
  configuration is a finite sequence of contiguous rigid sites (black dots).
  {\bf a)} On the square lattice, the size of this rigid cluster shrinks in
  time. {\bf b)} On the triangular lattice, the rigid cluster's size remains
  constant in time if the lattice is undiluted, but shrinks in time for any
  $p<1$, thus $p_c=1$ for this case. {\bf c)} On the undiluted 5n-lattice,
  rigid clusters grow in time. Therefore for this lattice there is a
  nontrivial value $p_c$ (depending on the type of dilution, i.e. bond or
  site) above which rigidity propagates forever. }
\label{fig:finiteclusters}
\end{figure}

Although no such time-reversal symmetry exists for DRP, we find that
$\beta^{dens}=\beta^{seed}$ also in this case.  Furthermore we find that DRP
belongs to the DP universality class, i.e. has exactly the same critical
indices. Thus, there is no separate universality class for directed rigidity
percolation as there is for \emph{undirected} rigidity percolation. This is
consistent with a recent conjecture~\cite{conjecture} according to which any
nonequilibrium process with a single absorbing state will belong to the same
universality class as DP.

We also studied the surface critical behavior~\cite{\DPW} of DRP, by means of
simulations in the presence of an absorbing boundary. In the DP case, the
presence of the absorbing wall is known to only modify the survival exponent
$\beta^{seed}$.  We find that this is also the case for DRP, and the new
exponent is also consistent with the one obtained for DP with a wall.

In section~\ref{sec:simulations} we present our numerical results for DRP on
directed lattices, with and without absorbing walls, and estimate the relevant
critical indices associated to the second-order transitions.
Section~\ref{sec:triangular} describes DRP on a directed triangular lattice.
This case has a first-order transition at $p=1$, and can be solved exactly for
$p\to 1$.

\section{Numerical simulations}
\label{sec:simulations}

In order to simulate DRP we store a binary variable per site, indicating
whether the given site is or not rigidly connected to the boundary at $t=0$.
We use the by now standard techniques of multispin-coding (MSC)~\cite{MSC},
which allow us to store 64 binary variables in an integer word, and also to
update all of them simultaneously. Since interactions are short-ranged in the
time direction, we only need to keep in memory a maximum of three consecutive
lines of the system and we do this by means of three linear arrays which are
reused periodically.

We have considered three different oriented lattices: square, triangular and
5n-lattice (see later). The first two of them display trivial behavior (i.e.
rigid only at $p=1$) and the third one presents a continuous DRP transition at
a non-trivial $p_c$.  In the first place we discuss DRP on a square lattice as
depicted in Fig.~\ref{fig:latexample}a. Each site at time $t$ has two
neighbors at time $t-1$. This is the minimum number of neighbors needed for
rigidity with $g=2$ and thus any amount of dilution is enough to impede
propagation of rigidity.  Therefore square lattices are not rigid at any
$p<1$. If $p=1$, rigidity propagates only if boundary conditions are
appropriate (e.g. rigid, or periodic, but not open). Any finite rigid cluster
of size $l$ shrinks to zero in $l$ time-steps, as shown in
Fig.~\ref{fig:finiteclusters}a. The same would happen on directed
$d$-dimensional hypercubic lattices if $g=d$.

We next consider the triangular lattice, oriented as shown in
Fig.~\ref{fig:latexample}b. Each site has three neighbors at earlier times.
Despite the number of neighbors being larger than the minimum required (two),
this lattice is also unable to propagate rigidity if diluted by any amount. To
see why this is so, consider Fig.~\ref{fig:finiteclusters}b, where one starts
from a finite cluster of rigid sites (black sites) at $t=0$. If the lattice is
undiluted ($p=1$), this cluster would just propagate unchanged in ``time''. If
the lattice is diluted by any amount, this rigid cluster would gradually
shrink and eventually disappear.  Thus for this lattice $p_c=1$, the same as
for the square lattice. In contrast, finite-size effects are expected to be
quite strong on the triangular lattice, since the lifetime of a finite rigid
cluster diverges as $p \to 1$, no matter its original size.  Also boundary
effects are different since now propagation of rigidity can exist without
periodic boundary conditions. We discuss this case in detail in
Sec.~\ref{sec:triangular}.

In order to have a nontrivial $p_c$ for DRP, we use triangular lattices
augmented with two further bonds per site. These extra bonds connect layers
$t$ and $t-2$, as shown in Fig.~\ref{fig:latexample}c. This makes a total of
five neighbors per site and we call this the 5n-lattice for simplicity. Now
consider what happens when starting from a finite cluster of rigid sites on an
undiluted 5n-lattice. As shown in Fig.~\ref{fig:finiteclusters}c, the size of
the rigid cluster \emph{expands} in time with a constant angle if $p=1$. Thus
there will be a nontrivial value $p_c$, above which rigidity propagates
forever. We find that the DRP transition is second-order on this lattice. For
comparison we also simulate DP on the square lattice.

We next discuss our numerical results for DRP on the 5n-lattice, and compare
them to DP on the square lattice.  We typically start our simulations from a
finite seed of contiguous rigid sites and let the system evolve for $10^5$
timesteps (or until all activity dies out) and measure the survival
probability $P_a^{seed}$, cluster width $w$ and average mass $M_{seed}$ as a
function of time.

\begin{figure}[htb]
  \centerline{ \psfig{figure=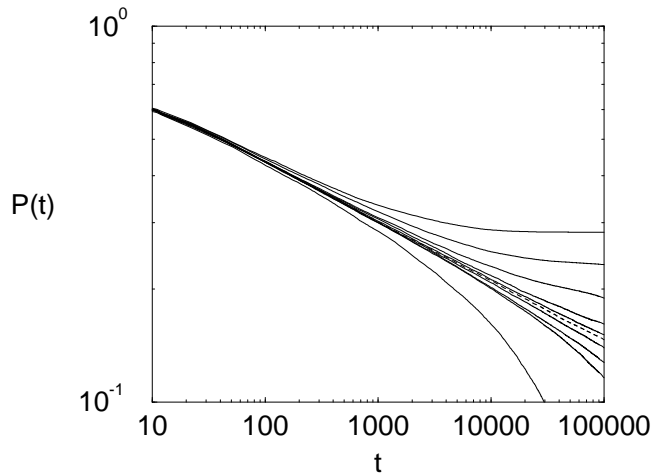,width=8cm,angle=270}}
\vskip 0.5cm
\caption{{} Survival probability for DRP clusters grown from finite seeds
  on site-diluted 5n-lattices with (from bottom to top) $p=0.70400, 0.70480,
  0.70490, 0.70500, 0.70505$ (dashed), $0.70510, 0.70520, 0.70550, 0.70600$
  and $0.70700$. Averages were taken over $10^5$ realizations on systems of
  width 3840 sites. }
\label{fig:p_t}
\end{figure}

In a first set of simulations we estimate the critical density $p_c$ for DRP
on site-diluted 5n-lattices, by measuring $P_a^{seed}(t)$ at different values
of $p$ and identifying the one for which the asymptotic behavior is closest to
a straight line in a log-log plot (Figure \ref{fig:p_t}). From these data we
estimate $p_c^{DRP}=0.70505 \pm 0.00005$.  In contrast to $P_a^{seed}$, which
shows appreciable curvature for off-critical values of $p$, the slopes of the
cluster mass $M_{seed}$ and the meandering width $w(t)$ in a similar log-log
graph show little variation when $p \neq p_c$. For DP on site-diluted square
lattices, we use the estimate~\cite{BG} $p_c=0.64470$.

\begin{figure}[htb] 
\vbox{
\centerline{{\bf a}\hskip -0.3cm\psfig{figure=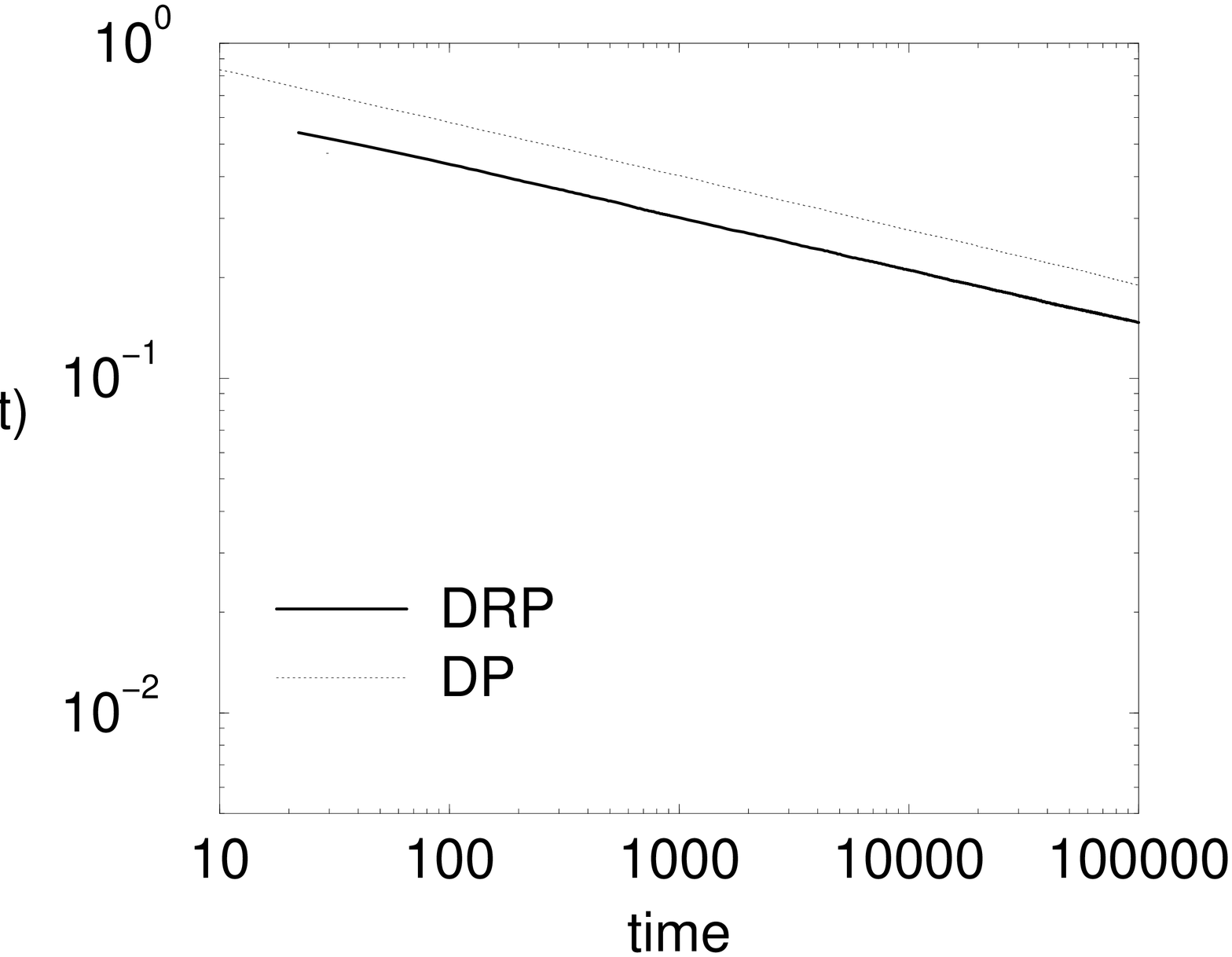,width=6cm,angle=270}}
\centerline{{\bf b}\hskip -0.3cm\psfig{figure=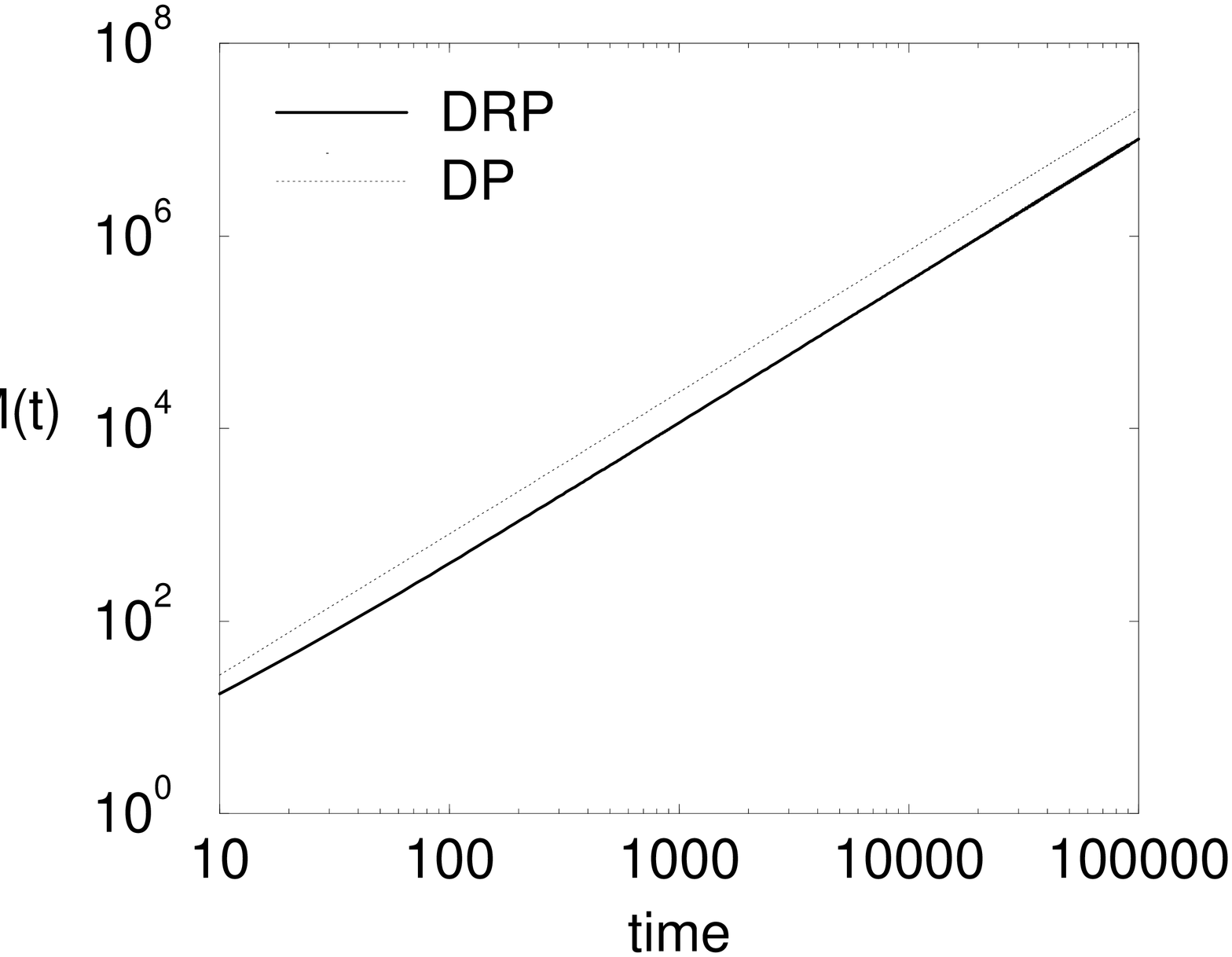,width=6cm,angle=270}}
\centerline{{\bf c}\hskip -0.3cm\psfig{figure=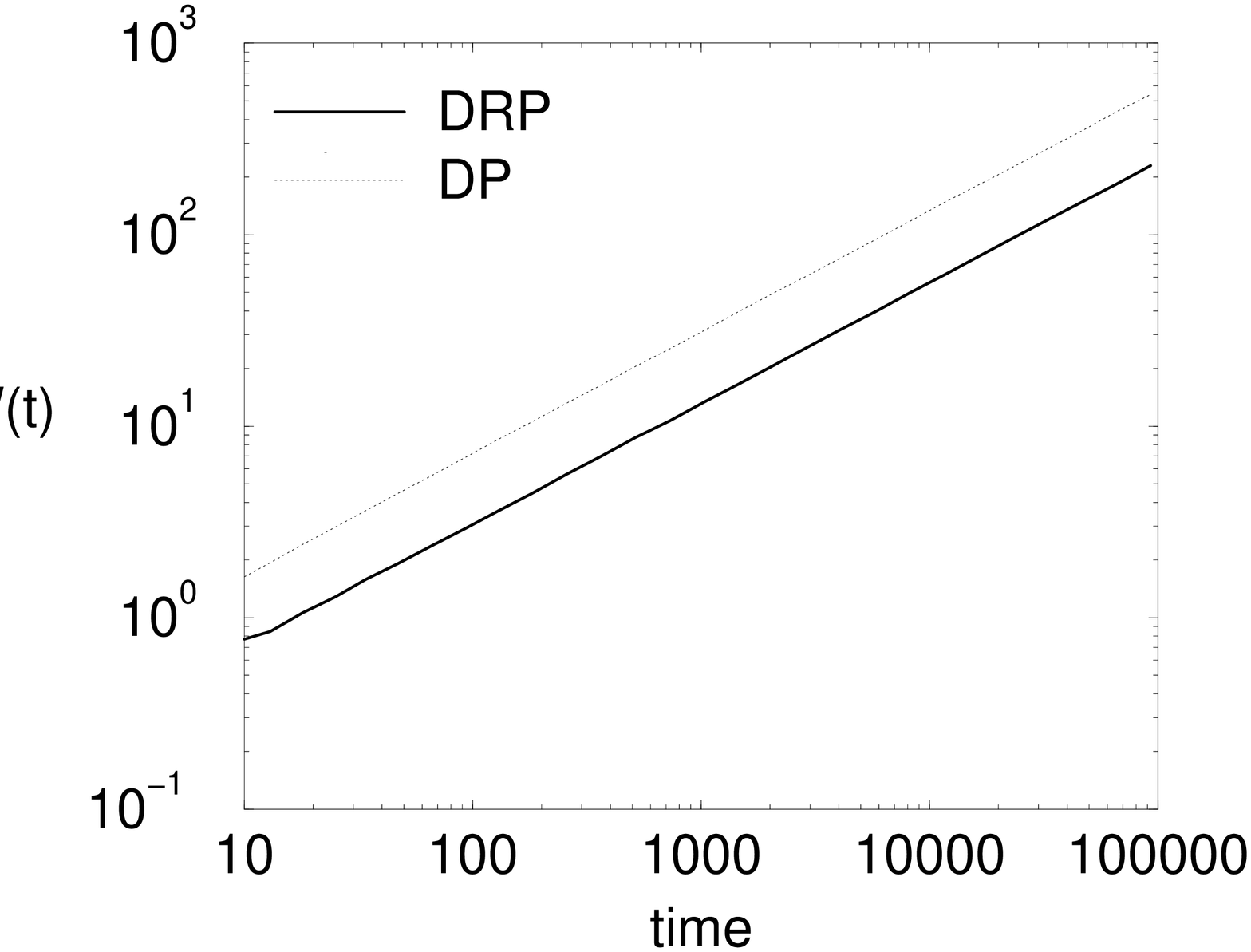,width=6cm,angle=270}}
}\vskip 0.5cm
\caption{{} Critical behavior obtained from finite-seeds at the second-order
  transition for DRP on site-diluted 5n-lattices and for DP on site-diluted
  square lattices.  {\bf a} Survival probability.  {\bf b} Cluster mass.
  {\bf c} Meandering width.  } 
\label{fig:nowall}
\end{figure}

Figure~\protect \ref{fig:nowall}a shows $P_{seed}(t)$ for DRP (5n-lattice)
and DP (square lattice) at their respective critical values. Assuming
power-law corrections to (\ref{eq:p-decay}), we fit $P_{seed}(p_c,t) = a
t^{-\delta}(1+b t^{-\omega})$ and find $\delta^{DRP} = 0.15 \pm 0.01$ and
$\delta^{DP}= 0.16 \pm 0.02$.

The cluster mass $M(t)$ and the meandering width $w(t)$ behave as shown in
Figures \ref{fig:nowall}b and c respectively.  From these data we estimate
$\tilde{\eta}^{DRP}=1.47 \pm 0.01$, $\tilde{\eta}^{DP}=1.47 \pm 0.01$,
$\chi^{DRP} = 0.633 \pm 0.005$ and $\chi^{DP} = 0.631 \pm 0.005$.

These estimates are consistent with the more precise values~\cite{LSMJ,BG}
$\delta^{DP}= 0.1594$, $\tilde{\eta}^{DP}= 1.4732$ and $\chi^{DP} = 0.6327$,
suggesting that DRP and DP are in the same universality class.

\begin{figure}[htb] 
\vbox{
\centerline{{\bf a}\hskip -0.3cm\psfig{figure=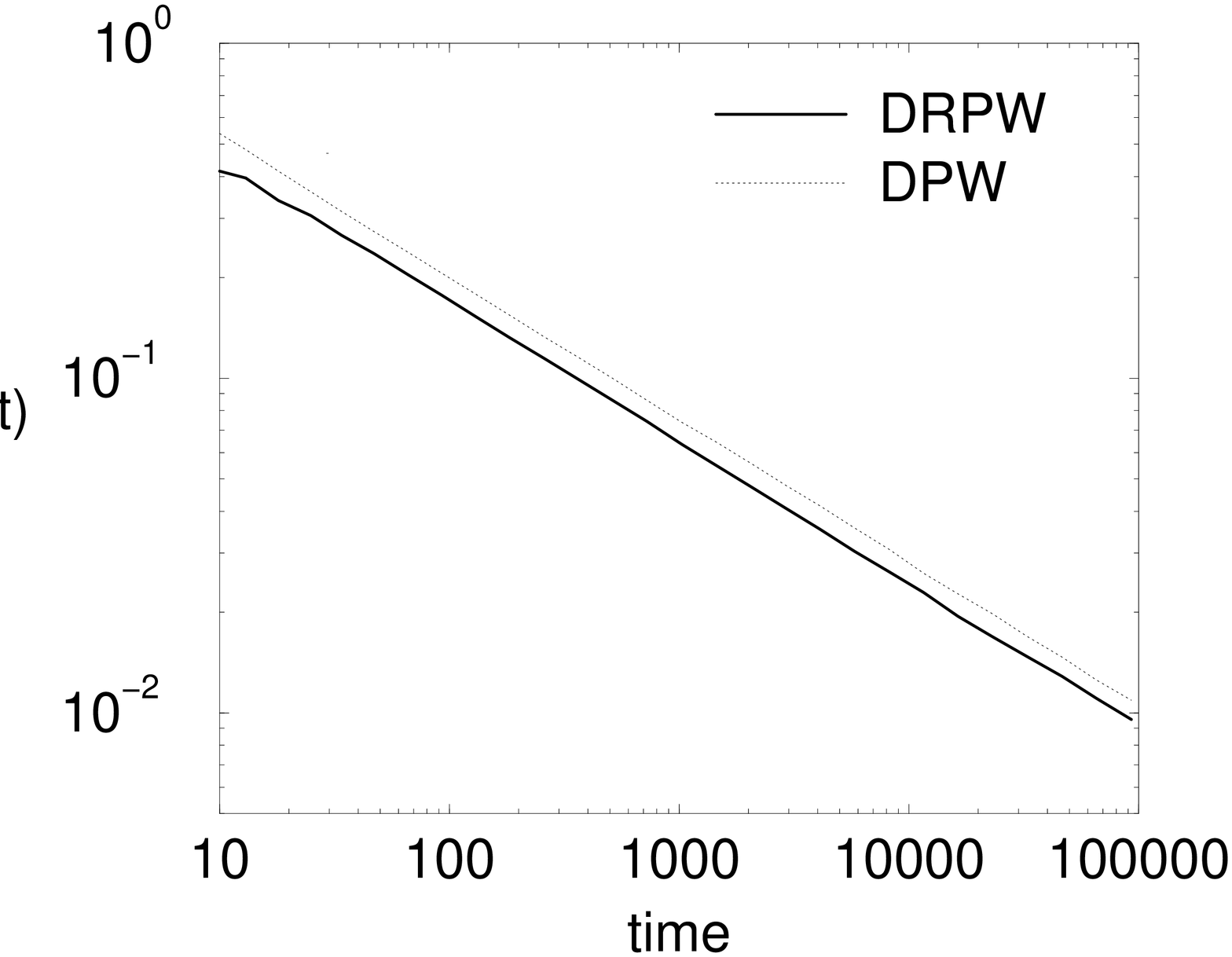,width=6cm,angle=270}}
\centerline{{\bf b}\hskip -0.3cm\psfig{figure=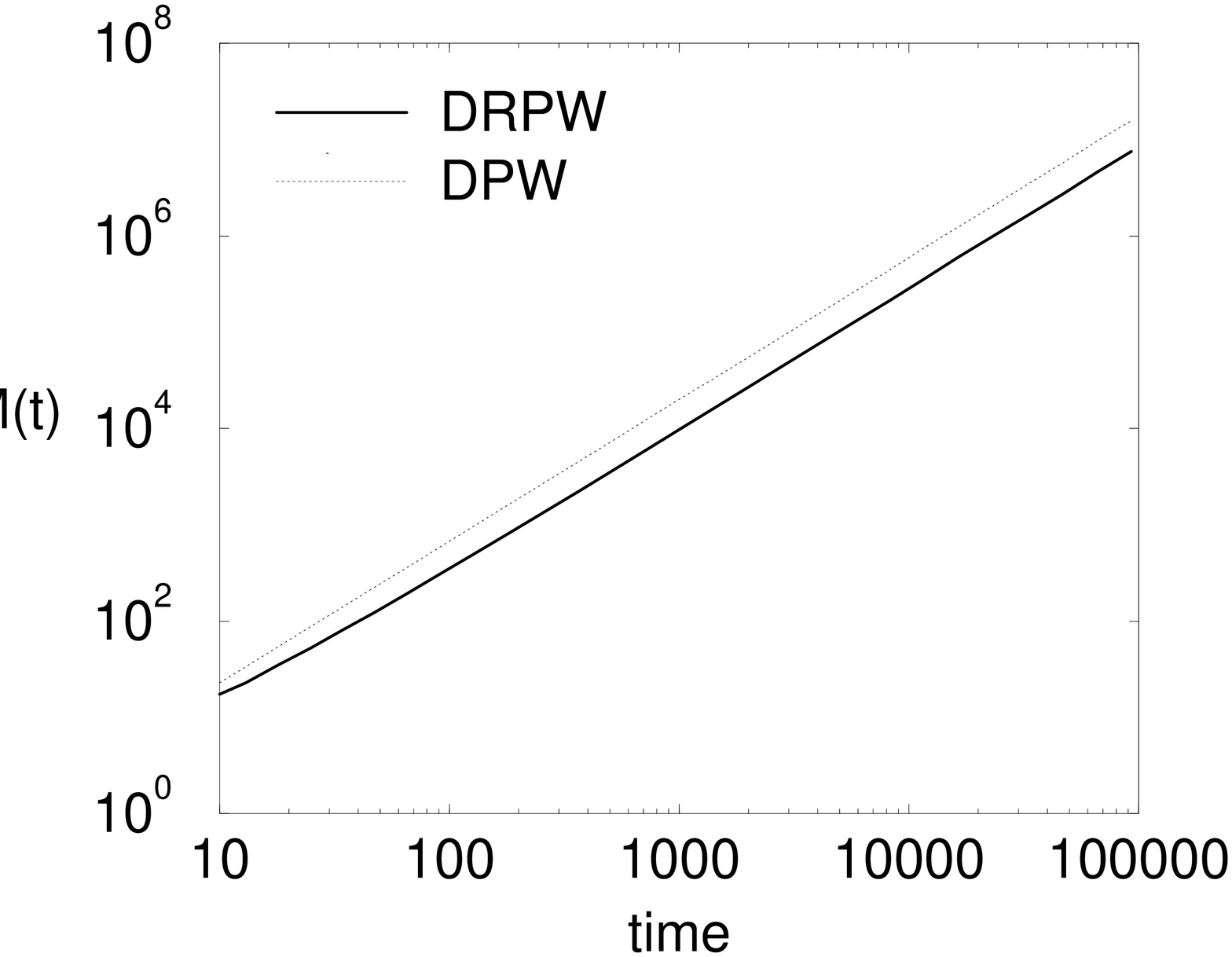,width=6cm,angle=270}}
\centerline{{\bf c}\hskip -0.3cm\psfig{figure=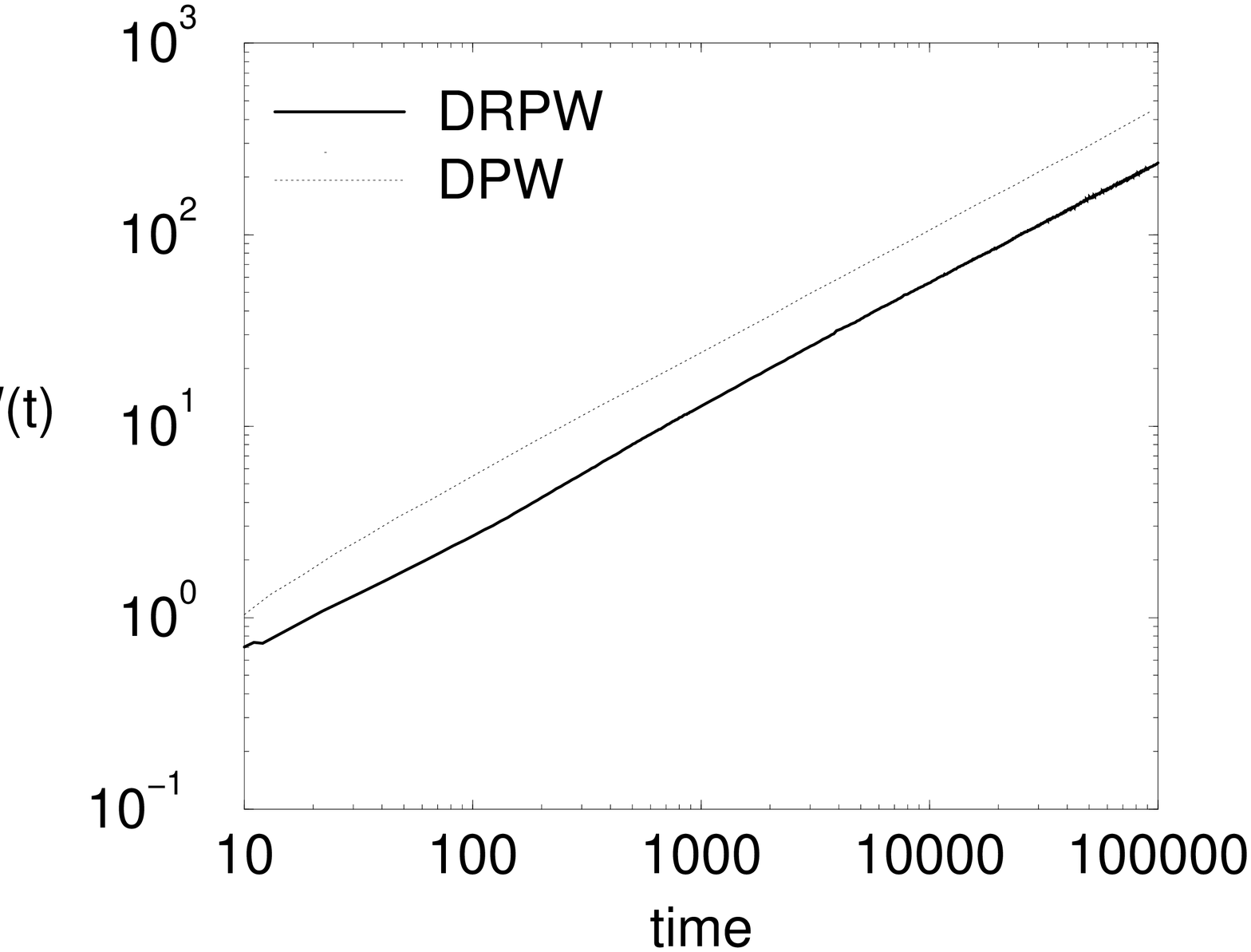,width=6cm,angle=270}}
\vskip 0.5cm
}\caption{{} Same as Fig.~\ref{fig:nowall}, but now finite clusters start near
  an absorbing wall at $x=0$. {\bf a} Survival probability.  {\bf b}
  Cluster mass.  {\bf c} Meandering width.  } 
\label{fig:wall}
\end{figure}

In order to further test whether DRP has the same critical behavior as DP, we
also studied DRP in the presence of an absorbing wall (DRPW). For DP with an
absorbing wall (DPW) it is known~\cite{\DPW} that the survival exponent
$\beta^{seed}$ is replaced by $\beta^{seed}_1$, while $\nu_{\parallel}$ and
$\nu_{\perp}$ remain unchanged. Therefore only $\delta$ is expected to change
due to the presence of the absorbing wall. Our results are displayed in
Fig.~\ref{fig:wall}, and from them we obtain $\delta^{DRPW}= 0.423 \pm 0.003$,
$\tilde{\eta}^{DRPW}= 1.48\pm 0.01$ and $\chi^{DRPW}= 0.62 \pm 0.02$. Notice
that $\delta^{DRPW}$, $\tilde{\eta}^{DRPW}$ and $-\chi^{DRPW}$ no longer add
up to one since $\beta^{dens}$ and $\beta^{seed}_1$ are independent exponents.
These results are entirely consistent with the values obtained for DPW by
other authors~\cite{LSMJ,FHL}.

\section{DRP on the triangular lattice}
\label{sec:triangular}

This case is marginal as already advanced, since any amount of dilution will
destroy rigidity and thus $p_c=1$, but on the other hand the lifetime of
finite clusters is not finite as on the square lattice, but diverges as $p\to
1$. As we show now, it is possible to obtain finite-size effects analytically
for $q=(1-p) << 1$.

\begin{figure}[htb]
\centerline{
{\bf a} \hskip -0.3cm
\psfig{figure=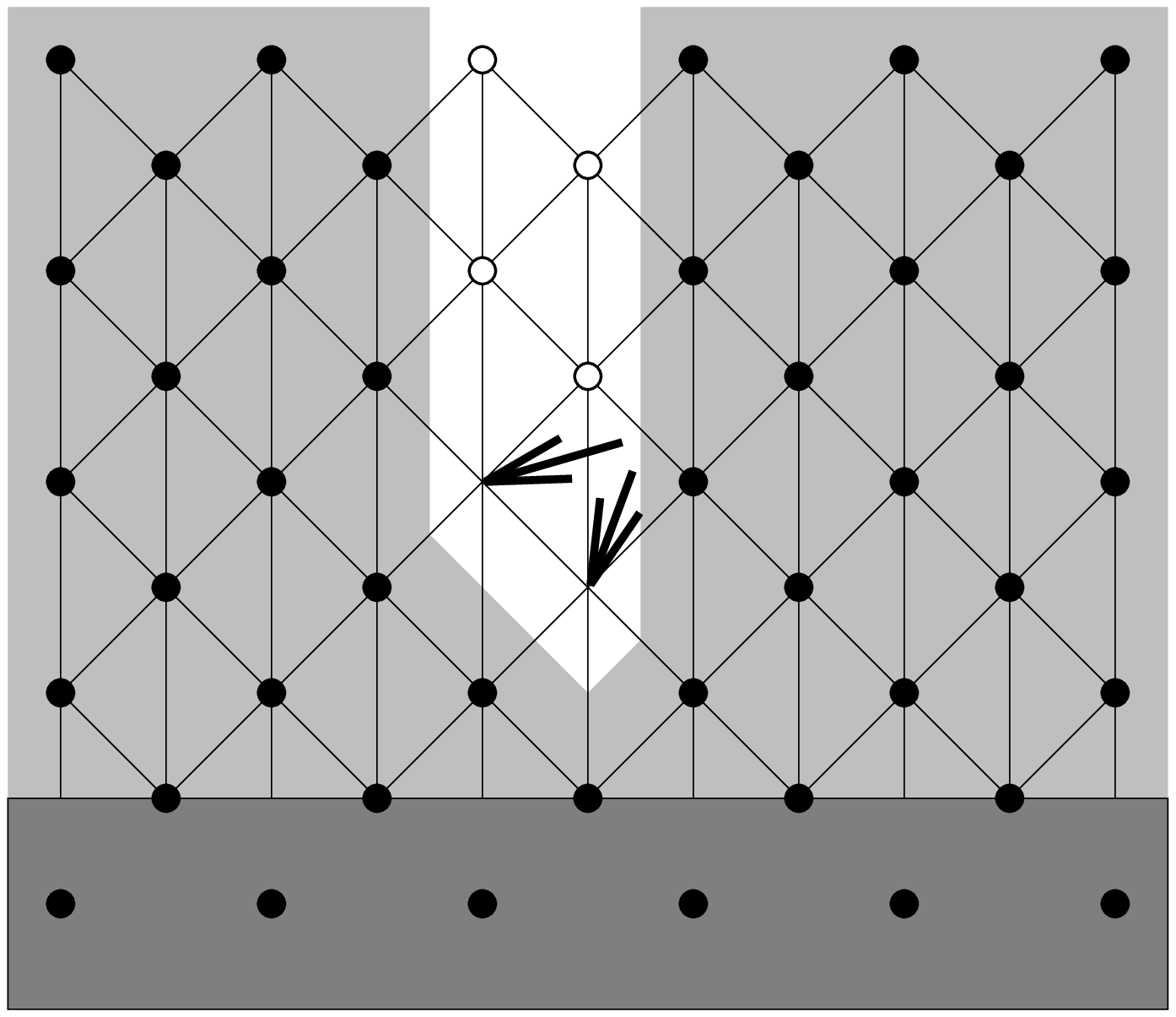,width=4cm,angle=270}
{\bf b} \hskip -0.3cm
\psfig{figure=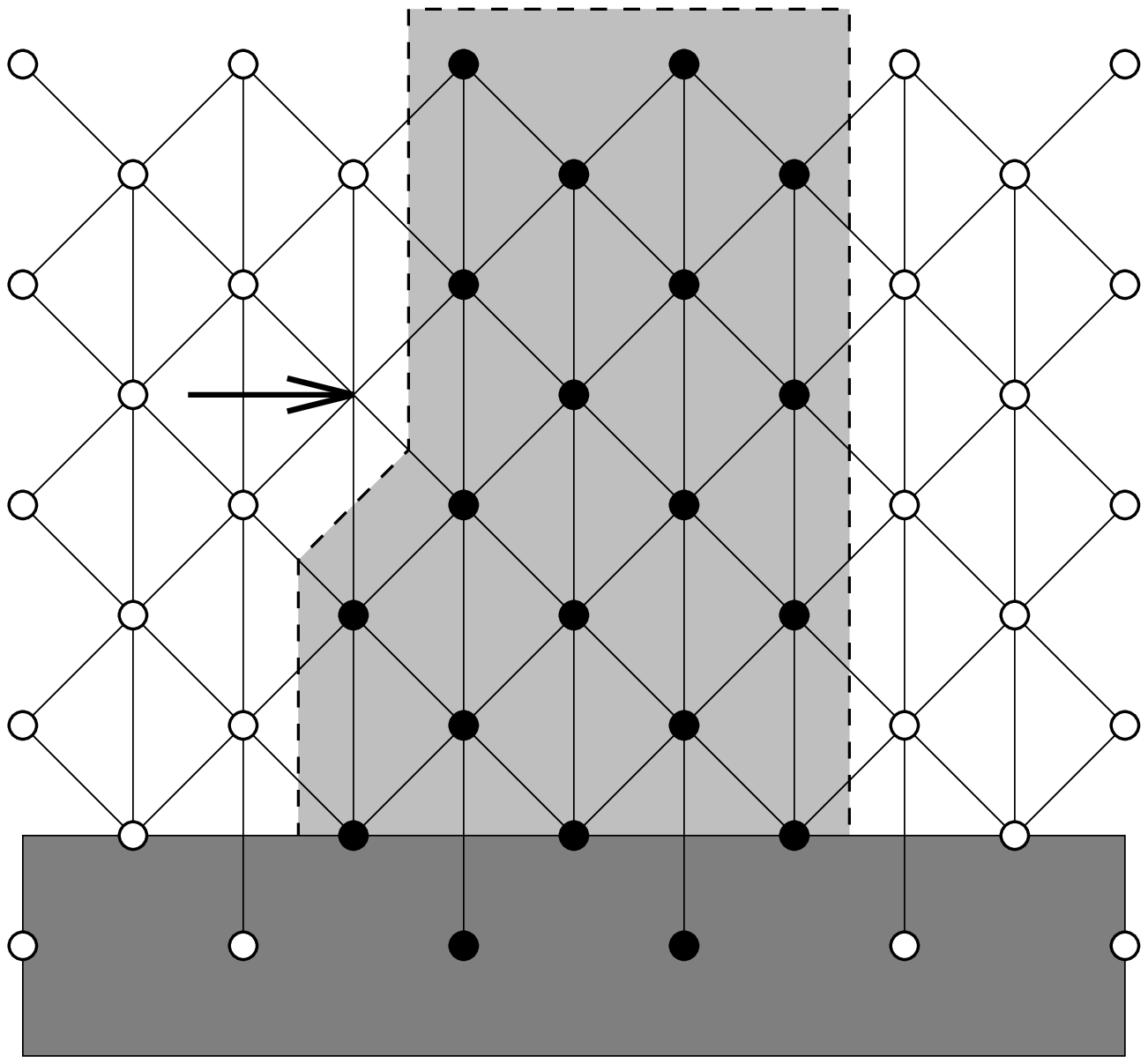,width=4cm,angle=270}
}
\vskip 0.5cm 
\caption{{} The directed triangular lattice presents a first-order transition
  at $p=1$, which can be described in simple terms. Arrows indicate absent
  sites, assuming site dilution. When $p<1$ {\bf a)} defects (pairs of
  adjacent non-rigid sites) nucleate at rate $\rho(p)$ per unit length, giving
  rise to non-rigid regions that {\bf b)} widen in time with velocity $v(p)$.}
\label{fig:defects}
\end{figure}

Assume one starts from a completely rigid boundary at $t=0$, on a triangular
lattice of infinite width (Fig.~\ref{fig:defects}a). Let $q=(1-p)<<1$ be the
dilution parameter. We do not need to specify for the moment whether we are
dealing with bond or site dilution. For short times all sites are rigidly
connected to the lower boundary, but soon some non-rigid sites, or
``defects'', will appear in the presence of either bond or site dilution.  The
smallest possible defect is a single non-rigid site, which happens with
probability $q$ per site on site-diluted lattices (one missing site) and with
probability $3 q^2p + q^3 \approx 3q^2$ per site on bond-diluted lattices (two
or three missing bonds). This single defect ``heals'' immediately since each
site above this one has three predecessors, but needs only two rigid ones in
order to be itself rigid.

A non-healing defect (in the following simply a defect) is created if two
sites connected by a diagonal bond are simultaneously non-rigid, as in
Fig.~\ref{fig:defects}a. All sites directly above these will have only one
rigid neighbor and thus fail to be rigid, creating a ``defect wall''.  Assume
that these paired defects are nucleated with density $\rho(p)$ per unit length
(we calculate $\rho$ later), and consider now the time evolution of the
resulting defect wall.

In the absence of dilution ($q=0$), the boundaries of a non-rigid region stay
unchanged in time (Fig.~\ref{fig:defects}a). Rigid sites directly on this
boundary have only two bonds (the minimum required number since $g=2$) to
rigid sites at earlier times.  If one of these boundary sites fails to be
rigid, all sites above it will also not be rigid.  In this case the rigid
boundary is displaced by one unit, as shown in Fig.~\ref{fig:defects}b.
Therefore, for small but nonzero $q$, the rigid wall in
Fig.~\ref{fig:defects}b moves rightwards with an average velocity $v =
\partial x / \partial t$ which equals the probability for a boundary site to
fail to be rigid.

\begin{figure}[htb]
  \centerline{ \psfig{figure=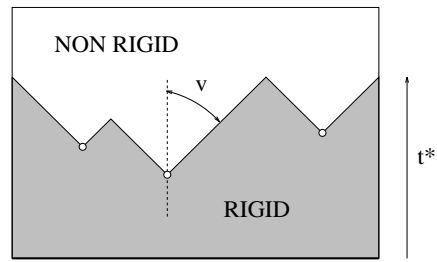,width=6cm,angle=270} }
\vskip 0.5cm
\caption{{}Defects (empty circles) nucleate at rate $\rho(p)$ per unit length, 
  and the resulting ``defect walls'' widen in time with velocity $v(p)$,
  eventually covering the entire system in a time of order $t^*$.}
\label{fig:coalescence}
\end{figure}

Neglecting fluctuations, we have a picture in which defects appear at a rate
$\rho(p)$ per unit length, giving rise to non-rigid regions which widen in
time with constant velocity $v(p)$. The system will become completely
non-rigid when all defect regions have coalesced, as depicted in
Fig.~\ref{fig:coalescence}.  This picture of the rigid-non-rigid transition is
related to the Polynuclear Growth model (PNG)\cite{png}, which has been
extensively studied in the area of crystal growth. For our discussion of DRP
we only need a few results which can be derived by means of simple arguments.

Assuming one knows the cone angle $v(p)$ and the defect density $\rho(p)$, it
is easy to calculate the density of rigid points $P(p,t)$ after $t$ timesteps
on a system of infinite width. A point $(x,t)$ will be rigidly connected to
the rigid boundary located at $t=0$ if it has not suffered the effect of any
defect. In other words, if no defect has nucleated inside a ``cone'' with
downwards opening angle $v$ and whose vertex sits at $(x,t)$.  Let
$\Omega=vt^2$ be the area of this cone. Since defects nucleate randomly in
space-time with density $\rho(p)$, their number inside any given area $\Omega$
is a Poisson-distributed random variable with average $\Omega \rho$. Thus
\begin{equation}
P(p,t) = e^{- \left( t/t^* \right )^2},
\end{equation}
where 
\begin{equation}
t^*= ( v \rho )^{-1/2}
\label{eq:tastsk}
\end{equation}
is a characteristic time for the disappearance of rigidity, on an infinitely
wide system.  Using similarly simple arguments it is easy to see that the mean
lifetime of a \emph{finite} rigid cluster diverges as $v^{-1}$ as $p \to 1$.

\begin{figure}[htb]
  \centerline{ \psfig{figure=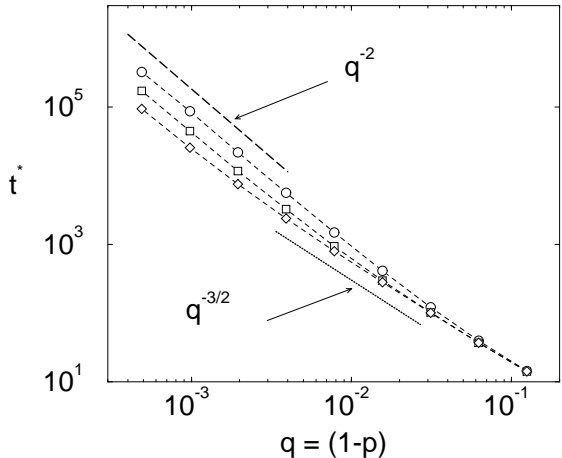,width=8cm,angle=270} }
\vskip 0.5cm
\caption{{} Mean time for the disappearance of rigidity, on
  site-diluted triangular lattices of width $w= 128$ (circles), $256$
  (squares) and $512$ (diamonds), as a function of $q=1-p$.}
\label{fig:lifetime}
\end{figure}

We now calculate $v(p)$ and $\rho(p)$, and compare the resulting prediction
for $t^*$ with our numerical results. Under site-dilution, the probability per
unit time for a rigid wall to be displaced by one unit is simply $v_{site}=q$.
If bonds are diluted instead, one gets $v_{bond}=1-p^2 \approx 2q$. In order
to calculate $\rho$, we notice that a pair of contiguous absent sites appears
with probability $\rho_{site}=2q^2$ per site and per unit time on site-diluted
lattices. On bond-diluted lattices on the other hand, creating such a pair
requires at least three missing bonds. Thus $\rho_{bond} \sim q^3$. Finally
one has (Eq.~(\ref{eq:tastsk})) $t^*_{site} \sim q^{-3/2}$ and $t^*_{bond}
\sim q^{-2}$.  Figure~\ref{fig:lifetime} shows $t^*$ as measured on
site-diluted lattices. These values are obtained by integrating in time the
density of rigid sites $P(p,t)$. For an intermediate range of $q$, it is found
that $t^* \sim q^{-3/2}$ as predicted for infinitely wide systems. 

On a system of a finite width $w$ with periodic boundary conditions, a
crossover to a width-dominated behavior is expected if $w << t^* v$,
equivalently if $w<< (v/\rho)^{1/2}$, whereupon
\begin{equation}
P(p,t)_{finite} = e^{- t/t^*_{finite} },
\end{equation}
and
\begin{equation}
t^*_{finite}= ( w \rho )^{-1}
\label{eq:tfinite}
\end{equation}
This regime corresponds to the defect-free area $\Omega$ becoming essentially
a rectangle of height $t$ and width $w$ instead of a triangle of height $t$
and base $vt$. According to equation (\ref{eq:tfinite}), one should expect
$t^*\sim q^{-2}$ for small $q$.

This regime is observed for $w=128$, but is less clear for larger values of
$w$. Observation of this crossover for wider systems is numerically difficult,
since it requires one to simulate very small $q$ values, which makes the mean
rigid times too large.

\section{Conclusions}

We considered directed rigidity percolation (DRP) with two degrees of freedom
per site on three different (1$+$1)-dimensional lattices. This problem is
equivalent to directed bootstrap percolation (DBP) with $m=2$. On the square
lattice, the system is only rigid at long times if $p=1$. On triangular
lattices a similar situation happens, but this case has a non-trivial behavior
for $p\to 1$, which we calculate analytically and confirm by numerical
simulation. The mean lifetime of rigidity on infinitely wide systems is found
to diverge when $p\to 1$ as $(1-p)^{-3/2}$ for site dilution, and as
$(1-p)^{-2}$ for bond dilution.  The mean lifetime of a finite rigid cluster
diverges on the other hand as $(1-p)^{-1}$ in both cases.

By augmenting the triangular lattice with two further bonds we define the
5n-lattice, which has a continuous transition at $p_c^{DRP}=0.70505 \pm
0.00005$ for site-dilution. We measure the critical indices associated with
the spreading of rigidity and find that the DRP transition belongs to the
directed percolation (DP) universality class, as a recent conjecture would
indicate. A similar numerical study of DRP with an absorbing wall gives
exponents equally consistent with those of DP. Thus, while (undirected)
rigidity percolation does not belong to the same universality class as usual
percolation, the introduction of directedness makes these two problems
essentially equivalent at their respective critical point, i.e. on large
scales. On (d$+$1) directed lattices, Bethe lattice calculations indicate that
the DRP transition becomes first-order for large $d$, while DP is always
second-order (with mean-field exponents above its upper critical dimension
$d_c=5$). We are presently extending this study to larger values of $d$.

\acknowledgements One of us (CM) wishes to thank K.~Lauritsen and
P.~Grassberger for useful discussions on DP. C.~M.  is supported by FAPERJ,
and M.~A.~M by CAPES, Brazil.


\end{document}